\documentclass[12pt]{article}
\usepackage{layout,amssymb,amsmath,epsfig}

\begin{document}
\title{\bf Anisotropic Strange Quintessence Stars in $f(R)$ Gravity}
\author{ G.
Abbas$^1$  \thanks{ghulamabbas@ciitsahiwal.edu.pk} M. Zubair$^2$
\thanks{mzubairkk@gmail.com; drmzubair@ciitlahore.edu.pk} G.
Mustafa$^1$ \thanks{gmjan@gmail.com} \\
$^1$ Department of Mathematics, COMSATS\\
Institute of Information Technology, Sahiwal, Pakistan.\\
$^2$ Department of Mathematics, COMSATS\\
Institute of Information Technology, Lahore, Pakistan.}

\date{}
\maketitle

\begin{abstract}
In this paper, we have formulated the new exact model of
quintessence anisotropic star in $f(R)$ theory of gravity. The
dynamical equations in $f(R)$ theory with the anisotropic fluid and
quintessence field have been solved by using Krori-Barua solution.
In this case, we have used the Starobinsky model of $f(R)$ gravity.
We have determined that all the obtained solutions are free from
central singularity and potentially stable. The observed values of
mass and radius of the different strange stars PSR J 1614-2230,
SAXJ1808.4-3658(SS1), 4U1820- 30, PSR J 1614-2230 have been used to
calculate the values of unknown constants in Krori and Barua metric.
The physical parameters like anisotropy, stability and redshift of
the stars have been investigated in detail.
\end{abstract}



\section{Introduction}

The discovery of cosmic acceleration is one the major advancements
in modern cosmology. The observation of type Ia supernovae (SNe Ia)
combined with observational probes of numerous mounting astronomical
evidences like the cosmic microwave background (CMB), large scale
structure surveys (LSS) and Wilkinson Microwave Anisotropy Probe
(WMAP) (Perlmutter et al.1999, Spergel et al.2007, Hawkins 2003,
Eisentein et al.2005) reveal that the cosmos at present is dominated
by exotic energy component named as dark energy (DE). The
investigation of current cosmic expansion and nature of DE has been
widespread among the scientists. For this purpose, numerous efforts
have been made based upon different strategies. These efforts can be
grouped in two categories: introducing new ingredients of DE to the
entire cosmic energy and modification of Einstein-Hilbert action to
obtain modified theories of gravity such as $f(R)$ (Nojiri and
Odintsov 2011) $f(\mathcal{T})$ (Ferraro and Fiorini 2007) where
$\mathcal{T}$ being the torsion, $f(R,T)$ (Harko et al.2011) where
$R$ and $T$ represent the scalar curvature and trace of the
energy-momentum tensor, $f(R,T,R^{\mu\nu}T_{\mu\nu})$ (Haghani 2013)
and Gauss-Bonnet gravities (Cognola 2006).

It has been the subject of great interest to study the models of
anisotropic stars solutions during the last decades (Herrera and
Santos 1997). Egeland (2007) investigated that the cosmological
constant would exist due to density of the vacuum, this is
consequence of modeling the mass and radius of the Neutron star. In
order to prove this fact Egeland used the relativistic equation of
hydrostatic equilibrium with fermion equation of state (EoS). As
$f(R)$ model with constant $R$ gives cosmological constant,
therefore motivated by this fact, we study the structure of strange
stars and concluded that $f(R)$ gravity with model
$f(R)=R+\lambda{R}^2$ (where $\lambda$ is constant) can describes
the class of some anisotropic compact strange stars for example
X-ray bruster 4$U$1820-30, X-ray pulsar Her X-1, Millisecond pulsar
SAX J 1808-3658 etc. in a very scientific way. During the recent
years, Dey et al.(1998), Usov (2004), Ruderman (1972), Mak and Harko
(2002, 2003) have studied the physical properties of strange stars
by using different approaches. Also, Herrera et al. (2008) have
examined all static spherically symmetric solution of Einstein field
equations and discussed the physical implication of these solutions.

The exact solutions have many applications in astronomy and
astrophysics. For examples these can explain the properties and
compositions of astrophysical objects. Mak and Harko (2004)
presented a class of exact solutions of Einstein field equations
with anisotropic source. The physical properties of these solutions
imply that matter density as well as tangential and radial pressure
are regular inside the compact star. Chaisi and Maharaj (2005)
established a mathematical algorithm which explain the solution of
the field equations for the anisotropic source. Rahaman et al.(2012)
apply the Krori-Barua (1975) solution to the charged strange compact
stars. Recently, Kalam et al.(2012) and Kalam et al.(2013) have
studied the models of the compact objects using the Krori-Barua
metric assumption. Hossein et al.(2012) discussed the anisotropic
star model in the presence of varying varying cosmological constant.
Bhar et al.(2015) investigated the higher dimensional compact star.
This work has been extended by Maurya et al.(2014) for the charged
anisotropic compact stars.

The study of compact stars such as relativistic massive objects have
been the subject of interest in GR as well as in modified theories
of gravity (Camenzind 2007, Abbas et al.2014, Abbas et al.2015a,
2015b, 2015c, 2015d, Sharif and Abbas 2013a, 2013b, Sharif and
Zubair 2013a, 2013b). Being highly dense these are small in size and
posses the extremely massive structure, and produce a strong
gravitational field. Recently, there has been growing interest to
study the compact stars in modified theories of gravity like $f(R)$
and $f(G)$. Particularly, Astashenok (2013), presented neutron stars
solutions for viable models of $f(G)$ gravity. Motivated by this
work, we study the compact stars solutions and their dynamical
stability for a viable model of $f(R)$ gravity. We shall find the
exact solutions for quintessence compact stars which are comparable
with observational data. Our plan in this work is as follows: In
Sec.\textbf{2}, we present basic equations of motion for $f(R)$
gravity. The analytic solution for the viable $f(G)$ model is
presented in Sec.\textbf{3}. Sec.\textbf{4} deals with the physical
analysis of the given system. The last section summaries the results
of the paper.

\section{Model of Anisotropic Quintessence star in $f(R)$ Gravity}

We assume that the manifold possesses a stationary and spherical symmetry for
which the metric can be written as
\begin{equation}\label{1}
ds^2=-e^{\mu(r)}dt^2+e^{\nu(r)}dr^2+r^2(d\theta^2+sin^2\theta{d\phi^2}),
\end{equation}
where $\nu=Ar^2$, $\mu=Br^2+C$ Krori and Barua (1975), $A$, $B$ and
$C$ are arbitrary constant
to be evaluated by using some physical conditions.\\
$f(R)$ gravity is defined by the action
\begin{equation}\label{2}
\mathcal{\mathcal{I}}=\int{dx^4\sqrt{-g}\left(f(R)+\mathcal{L}_{(matter)}\right)},
\end{equation}
where $\kappa^2=1,f(R)$ is generic function of scalar curvature and
$\mathcal{L}_{(matter)}$ determines the role of matter contents. Variation of
action (\ref{1}) with respect to $g_{\mu\nu}$ yields the field equations
\begin{equation}\label{3}
G_{{\mu}{\nu}}=T_{{\mu}{\nu}}^{(curv)}+\mathcal{T}_{{\mu}{\nu}},
\end{equation}
where
$\mathcal{T}_{{\mu}{\nu}}=T_{{\mu}{\nu}}^{matter}+\mathcal{Q}_{{\mu}{\nu}}$,
which contains of ordinary matter contribution and quintessence
filed defined by the parameter $-1<\omega_q<-1/3$ (Bhar 2015).
$\mathcal{T}_{{\mu}{\nu}}$ is scaled by a factor of
$\frac{1}{f_R(R)}$ and $T_{{\mu}{\nu}}^{(curv)}$ denotes the
contribution that arises from the curvature to the effective
stress-energy tensor given by
\begin{equation}\label{4}
T_{{\mu}{\nu}}^{(curv)}=\frac{1}{f_R(R)}\left[\frac{1}{2}g_{\mu\nu}(f(R)-Rf_R(R))
+f_R(R)^{;{\alpha}{\beta}}(g_{\mu\alpha}g_{\nu\beta}-g_{\mu\nu}g_{\alpha\beta})\right],
\end{equation}
where $f_R(R)$, denotes derivative with respect to the Ricci scalar
$R$.

The components of $\mathcal{Q}^\mu_\nu$ are defined as (Bhar 2015)
$\mathcal{Q}^t_t=\mathcal{Q}^r_r=-\rho_q$ and
$\mathcal{Q}^\theta_\theta=\mathcal{Q}^\phi_\phi=\frac{1}{2}(3\omega_q+1)\rho_q$.
To obtain some particular strange star models, we assume the
anisotropic fluid in the interior of compact object and it is
defined by
\begin{equation}\label{5}
T^m_{\alpha\beta}=(\rho+p_t)u_\alpha{u}_\beta-p_tg_{\alpha\beta}+(p_r-p_t)v_\alpha{v}_\beta,
\end{equation}
where $u_\alpha=e^\frac{\mu}{2}\delta^0_\alpha$,
$v_\alpha=e^\frac{\nu}{2}\delta^0_\alpha$, $\rho$, $p_r$ and $p_t$
correspond to energy density, radial and transverse pressures,
respectively. The modified field equations corresponding to
spacetime (\ref{4}) lead to
\begin{eqnarray}\nonumber
&&\rho+\rho_q=-e^{-\nu}f_{RRR}+e^{-\nu}\left(\frac{\nu'}{2}-\frac{2}{r}\right)f_{RR}
+\frac{e^{-\nu}}{r^2}\left(\frac{\mu''r^2}{2}+\frac{\mu'^2r^2}{4}
-\frac{\mu'\nu'r^2}{4}\right.\\\label{6}&+&\left.\mu'r\right)f_R-\frac{1}{2}f,\\\nonumber
&&p_r+\rho_q=
e^{-\nu}\left(\frac{\mu'}{2}+\frac{2}{r}\right)f_{RR}-\frac{e^{-\nu}}{r^2}
\left(\frac{\mu''r^2}{2}+\frac{\mu'^2r^2}{4}-\frac{\mu'\nu'r^2}{4}
-\nu'r\right)f_R\\\label{7}&+&\frac{1}{2}f,\\\nonumber
&&p_t+\frac{1}{2}(3\omega_q+1)\rho_q=-e^{-\nu}f_{RRR}+e^{-\nu}\left(\frac{\mu'}{2}-\frac{\nu'}{2}
+\frac{1}{r}\right)f_{RR}-\frac{e^{-\nu}}{r^2}\left(\frac{\mu'r}{2}
\right.\\\label{8}&-&\left.\frac{\nu'r}{2}-e^{\nu}+1\right)f_R+\frac{1}{2}f.
\end{eqnarray}
Herein, we choose the Starobinsky model of the form
\begin{equation}\label{9}
f(R)=R+\lambda{R}^2,
\end{equation}
where $\lambda$ is an arbitrary constant. One can set $\lambda=0$ to
find the results in GR. Herein, we set $\lambda=2km^2$.

For this model the equations (\ref{6})-(\ref{8}) become
\begin{eqnarray}\nonumber
&&\rho+\rho_q=\frac{e^{-2\nu}}{8r^4}\{r^4\lambda\mu'^4-2r^3\lambda\mu'^3
(-4+r\nu')+r^2\lambda\mu'^2(16+8r\nu'-11r^2\nu'^2+4r^2\mu''\\\nonumber&+&8r^2\nu'')+4r^2\lambda
\mu'(-16r\nu'^2+3r^2\nu'^3+\nu'(-4+9r^2\mu''-7r^2\nu'')+2r(-2\mu''\\\nonumber&+&6\nu''
-2r\mu'''+r\nu'''))+4(-2e^\nu{r}^2+2e^{2\nu}r^2-20\lambda+24e^\nu
\lambda-4e^{2\nu}\lambda\\\nonumber&+&12r^3\lambda\nu'^3-3r^4\lambda\mu''^2-r^2\lambda
\nu'^2(12+11r^2\mu'')+8r^2\lambda\nu''+8r^4\lambda\mu''\mu''-16\\\nonumber&\times&r^3\lambda\mu'''+
2r\nu'(e^\nu{r}^2-8\lambda+16r^2\lambda\mu''-14r^2\lambda\nu''+6r^3\lambda\mu''')+
8r^3\lambda\nu'''\\\label{10}&-&4r^4\lambda\mu^{(iv)})\},\\\nonumber
&&p_r-\rho_q=\frac{e^{-2\nu}}{8r^4}\{-r^4\lambda\mu'^4-2r^4\lambda\mu'^3\nu'+r^3
\lambda\mu'^2(-24\nu'+3r\nu'^2+4r(\mu''-\nu''))\\\nonumber&-&4(-2e^\nu{r}^2+2e^{2\nu}r^2+28\lambda
-24e^\nu\lambda-4e^{2\nu}\lambda-12r^2\lambda\nu'^2-16r^2\lambda\mu''\\\nonumber&+&
8r^3\lambda\nu'\mu''+r^4\lambda\mu''^2+16r^2\lambda\nu''-8r^3
\lambda\mu'''+8r\mu'(e^\nu{r}^2-8\lambda+3r^2\lambda\nu'^2\\\label{11}&+&6r^2
\lambda\mu''-r\lambda\nu'(8+r^2\mu'')-4r^2\lambda\nu''+r^3\lambda\mu''')\},\\\nonumber
&&p_t+\frac{1}{2}(3\omega_q+1)\rho_q=\frac{e^{-\nu}}{8r^4}\{r^4\lambda\mu'^4+2r^3\lambda\mu'^3(2-3r\nu')+
r^2\mu'^2(-32r\lambda\nu'\\\nonumber&+&17r^2\lambda\nu'^2+2(e^\nu{r}^2-8\lambda+6r^2
\lambda\mu''-6r^2\lambda\nu''))-2r\mu'(-38r^2\lambda\nu'^2+6r^3
\lambda\nu'^3\\\nonumber&+&r\nu'(e^\nu{r}^2-24\lambda+28r^2\lambda\mu''-14r^2
\lambda\nu'')-2(e^\nu{r}^2-4\lambda+12e^nu\lambda+10r^2\lambda\mu''\\\nonumber&-&
14r^2\lambda\nu''+6r^3\lambda\mu'''-2r^3\lambda\nu'''))-4(12r^3
\lambda\nu'^3-11r^4\lambda\mu''\nu'^2-5r^4\lambda\mu''^2\\\nonumber&+&\mu''(-e^\nu
r^4+8r^2\lambda+8r^4\lambda\nu'')+r\nu(e^\nu{r}^2-28\lambda+12e^\nu
\lambda+28r^2\lambda\mu''-28r^2\lambda\nu''\\\label{12}&+&12r^3\lambda\mu''')+4\lambda
(-7+6e^\nu+e^{2\nu}-3r^3\mu'''+2r^3\nu'''-r^4\mu^{(iv)}))\}.
\end{eqnarray}

Using the relations of $\mu$ and $\nu$, we find the following
relations
\begin{eqnarray}\nonumber
&&\rho+\rho_q=\frac{1}{r^4}e^{-2Ar^2}\{e^{2Ar^2}(r^2-2\lambda)+2(-5-3B^2r^4+6B^3r^6
+B^4r^8\\\nonumber&+&12A^3r^6(2+Br^2)-A^2r^4(40+68Br^2+11B^2r^4)+A(-4r^2
+48Br^4\\\label{13}&+&26B^2r^6
-2B^3r^8))\lambda+e^{Ar^2}(-r^2+2Ar^4+12\lambda)\},\\\nonumber
&&p_r-\rho_q=\frac{1}{r^4}e^{-2Ar^2}\{-e^(2Ar^2)(r^2-2\lambda)+2(-7+11B^2r^4+
2B^3r^6\\\nonumber&-&B^4r^8+3A^2r^4(2+Br^2)^2-2Ar^2(4+16Br^2+9B^2r^4+
B^3r^6))\lambda+e^{Ar^2}(r^2\\\label{14}&+&2Br^4+12\lambda)\},\\\nonumber
&&p_t+\frac{1}{2}(3\omega_q+1)=\frac{1}{r^4}e^{-2Ar^2}\{-2(-7+6e^{Ar^2}+e^{2Ar^2})\lambda+16B^3r^6\lambda
\\\nonumber&+&2B^4r^8\lambda-24A^3r^6(2+Br^2)\lambda+2A^2r^4(28+74Br^2+17B^2r^4)\lambda+
B^2r^4\\\nonumber&\times&(e^{Ar^2}r^2+22\lambda)+2Br^2(-6\lambda+e^{Ar^2}(r^2+6\lambda))-
Ar^2(4(-7+19Br^2\\\label{15}&+&25B^2r^4+3B^3r^6)\lambda+e^{Ar^2}(r^2+Br^4+12\lambda))\}.
\end{eqnarray}
We have four unknown functions $\rho, p_r, p_t, \rho_q$, and three
Eqs.(\ref{13})-(\ref{15}). To find the explicit relations of these
parameters, we assume a relation between radial pressure and energy
density of the form
\begin{equation}\label{15a}
p_r=\alpha\rho, \quad 0<\alpha<1,
\end{equation}
where $\alpha$ plays the role of equation of state parameter. This
relation is the particular form of equation of state, the general
form of this equation has been presented by Herrera and Barreto
(2013).

After some manipulations and using Eq.(\ref{15a}), we get the
expressions of $\rho, p_r, p_t, \rho_q$ in the following form
\begin{eqnarray}\nonumber
\rho&=&\frac{1}{1+\alpha}\left\{\frac{1}{r^4}e^{-2Ar^2}(8
(-3-3Ar^2+(-7A^2+4AB+2B^2)r^4+2(3
A^3\right.\\\nonumber&-&\left.7A^2B+AB^2+B^3)\times{r}^6+A(A-B)B(3A+B)r^8)
\lambda+2e^{Ar^2}((A\right.\\\label{15b}&+&\left.B)r^4+12\lambda))\right\},\\\nonumber
\rho_q&=&\frac{1}{r^4(1+\alpha)}e^{-2Ar^2}\left\{e^{2Ar^2}(1+\alpha)
(r^2-2\lambda)+2(7-5\alpha+12A^3r^6(2+Br^2)\alpha\right.\\\nonumber&+&\left.B^2r^4
(-11-3\alpha+Br^2(-2+6\alpha+Br^2(1+\alpha)))-A^2r^4(12+40\alpha+Br^2\right.\\\nonumber&\times&\left.(12+
68\alpha+Br^2(3+11\alpha)))+2Ar^2(4-2\alpha+Br^2(8(2+3\alpha)+Br^2(9
\right.\\\nonumber&-&\left.Br^2(-1+\alpha)+13\alpha))))\lambda-e^{Ar^2}(r^2(1+\alpha)+2r^4(B-A\alpha)
\right.\\\label{15c}&-&\left.12(-1+\alpha)\lambda)\right\},
\\\nonumber
p_r&=&\frac{\alpha}{1+\alpha}\left\{\frac{1}{r^4}e^{-2Ar^2}(8
(-3-3Ar^2+(-7A^2+4AB+2B^2)r^4+2(3
A^3\right.\\\nonumber&-&\left.7A^2B+AB^2+B^3)\times{r}^6+A(A-B)B(3A+B)r^8)
\lambda+2e^{Ar^2}((A\right.\\\label{15d}&+&\left.B)r^4+12\lambda))\right\},\\\nonumber
p_t&=&\frac{1}{2r^4(1+\alpha)}e^{-2Ar^2}\left\{-e^{2Ar^2}(1+\alpha)
(\lambda(2-6\omega
_q)+r^2(1+3\omega_q))+e^{Ar^2}\right.\\\nonumber&\times&\left.(-2(A-B)Br^6
(1+\alpha )+r^2(1+\alpha )(1+24(-A+B)\lambda+3\omega
_q)-12\lambda\right.\\\nonumber&\times&\left. (1+3\alpha+3(-1+\alpha
)\omega _q)-2r^4(A+A\alpha(2+3\omega
_q)-B(3+2\alpha+3\omega_q)))\right.\\\nonumber&-&\left.2\lambda(-7-19\alpha+12Br^2(1+\alpha
)+3(7-5\alpha)\omega_q+B^4r^8(1+\alpha)(-1+3\omega
_q)\right.\\\nonumber&+&\left.2B^3r^6(-9-5
\alpha+(-3+9\alpha)\omega_q)-B^2r^4(33+25\alpha+(33+9\alpha)\omega_q)\right.\\\nonumber&+&\left.12
A^3r^6(2+Br^2)(2+3\alpha(1+\omega_q))+2Ar^2(-2(5+8\alpha+3(-2+\alpha)\omega
_q)\right.\\\nonumber&+&\left.B^3r^6(7+5\alpha-3(-1+\alpha)\omega_q)+2Br^2(27+31\alpha+12(2+3
\alpha)\omega_q)+B^2r^4\right.\\\nonumber&\times&\left.(59+63\alpha+3(9+13\alpha)\omega_q))-A^2r^4(4(17+9
\omega_q+6\alpha(4+5\omega_q))+B^2r^4\right.\\\nonumber&\times&\left.(34+\alpha(45+33\omega_q))+Br^2(r^2(3+9
\omega_q)+4(40+54\alpha+(9+51\alpha)\omega_q))))\right\}.\\\label{15e}
\end{eqnarray}
The equation of state (EoS) parameters corresponding to radial and
transverse directions can be obtained as
\begin{eqnarray}\label{16}
\omega_r&=&\alpha,\\\nonumber \omega_t&=&\left\{-e^{2 A
r^2}(1+\alpha)(\lambda(2-6\omega_q)+r^2(1+3\omega_q))+e^{Ar^2}
(-2(A-B)Br^6(1\right.\\\nonumber&+&\left.\alpha)+r^2(1+\alpha)(1+24(-A+B)\lambda+3\omega_q)
-12\lambda(1+3\alpha+3(-1+\alpha)\omega_q)\right.\\\nonumber&-&\left.2r^4(A+A\alpha(2+3\omega_q)
-B(3+2\alpha+3\omega_q)))-2\lambda(-7-19\alpha+12Br^2(1\right.\\\nonumber&+&\left.\alpha)
+3(7-5\alpha)\omega_q+B^4r^8(1+\alpha)(-1+3\omega_q)+2B^3r^6(-9
-5\alpha+(-3+9\alpha)\right.\\\nonumber&\times&\left.\omega_q)-B^2r^4(33+25\alpha+(33+9\alpha)\omega_q)
+12A^3r^6(2+Br^2)(2+3\alpha(1+\omega_q))\right.\\\nonumber&+&\left.2Ar^2(-2
(5+8\alpha+3(-2+\alpha)\omega_q)+B^3r^6(7+5\alpha-3(-1+\alpha)\omega_q)
+2Br^2\right.\\\nonumber&\times&\left.(27+31\alpha+12(2+3\alpha)\omega_q)+B^2r^4(59+63\alpha+3(9+13\alpha)
\omega_q))-A^2r^4(4\right.\\\nonumber&\times&\left.(17+9\omega_q+6\alpha(4+5\omega_q))+B^2r^4(34+\alpha(45
+33\omega_q))+Br^2(r^2(3+9\omega_q)\right.\\\nonumber&+&\left.4(40+54\alpha+(9+51\alpha)\omega_q))))\right\}
/\left\{2(8(-3-3Ar^2+(-7A^2+4AB+2B^2)r^4\right.\\\nonumber&+&\left.2(3A^3-7A^2B+AB^2+B^3)r^6+A(A-B)B(3A+B)r^8)
\lambda+2e^{Ar^2}((A\right.\\\label{17}&+&\left.B)r^4+12\lambda))\right\}.
\end{eqnarray}
We show the evolution of energy density $\rho$, radial pressure
$p_r$, tangential pressure $p_t$ and quintessence field $\rho_q$ for
the strange star candidates Her X-1, SAX J 1808.4-3658 and 4U
1820-30 in Figures \textbf{1-4}.

\begin{figure}
\centering \epsfig{file=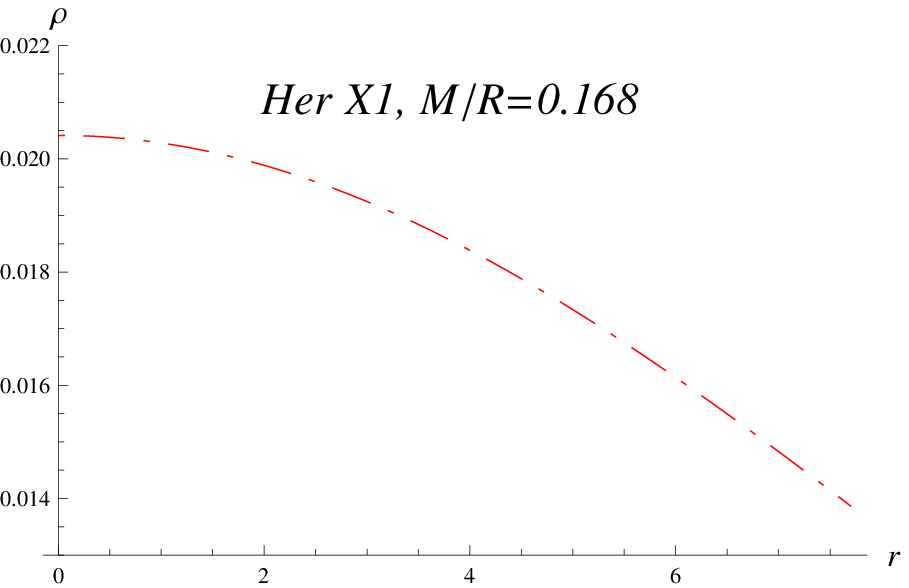, width=.34\linewidth,
height=1.4in}\epsfig{file=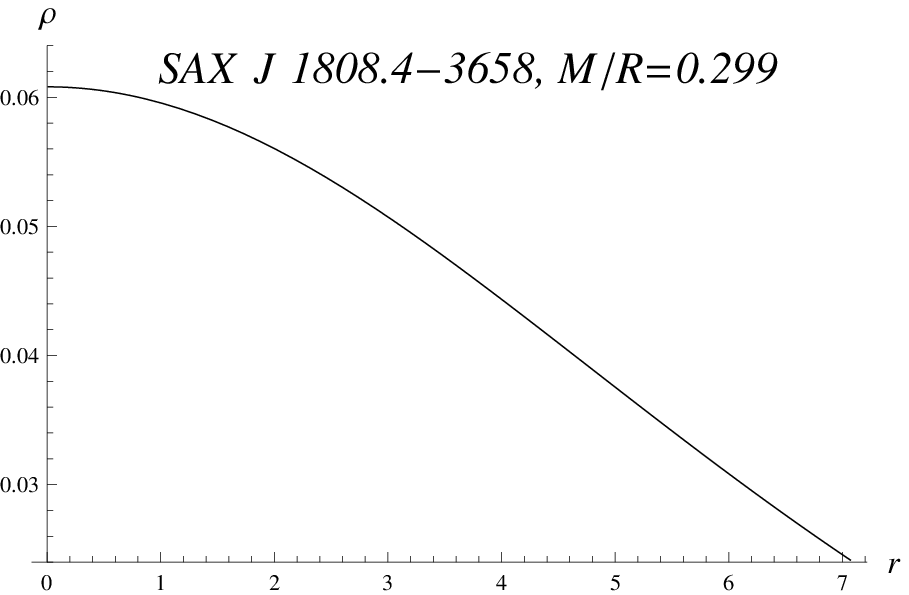, width=.36\linewidth,
height=1.4in}\epsfig{file=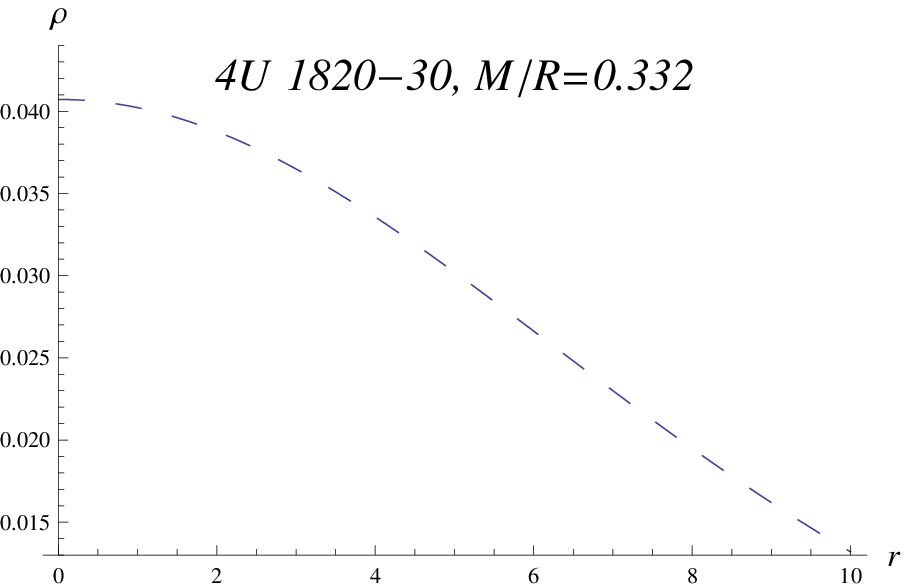, width=.34\linewidth,
height=1.4in}\caption{Evolution of energy density $\rho$ versus
$r(km)$ at the stellar interior of strange star candidates. Herein,
we set $\lambda=2km^2$, $\alpha=.01$ and $\omega_q=0.4$.}
\end{figure}
\begin{figure}
\centering \epsfig{file=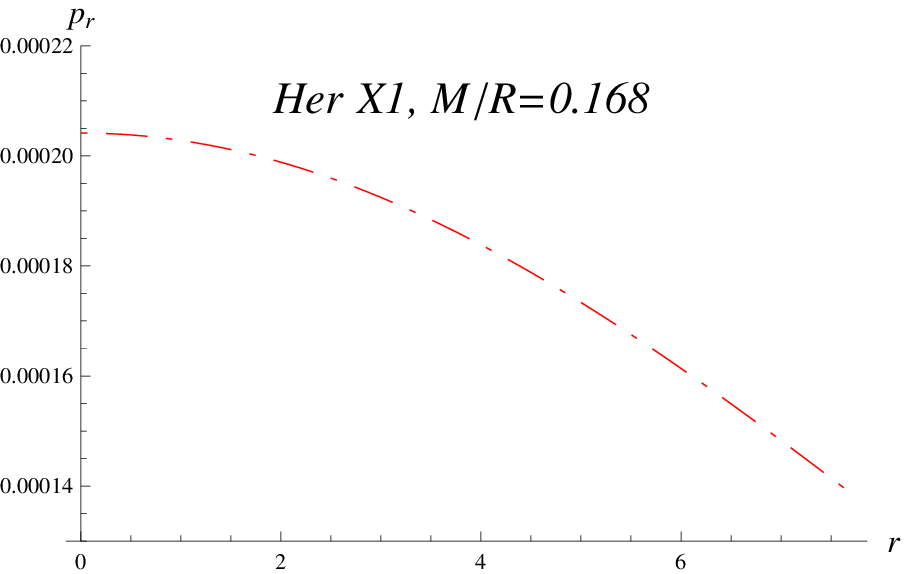, width=.34\linewidth,
height=1.4in}\epsfig{file=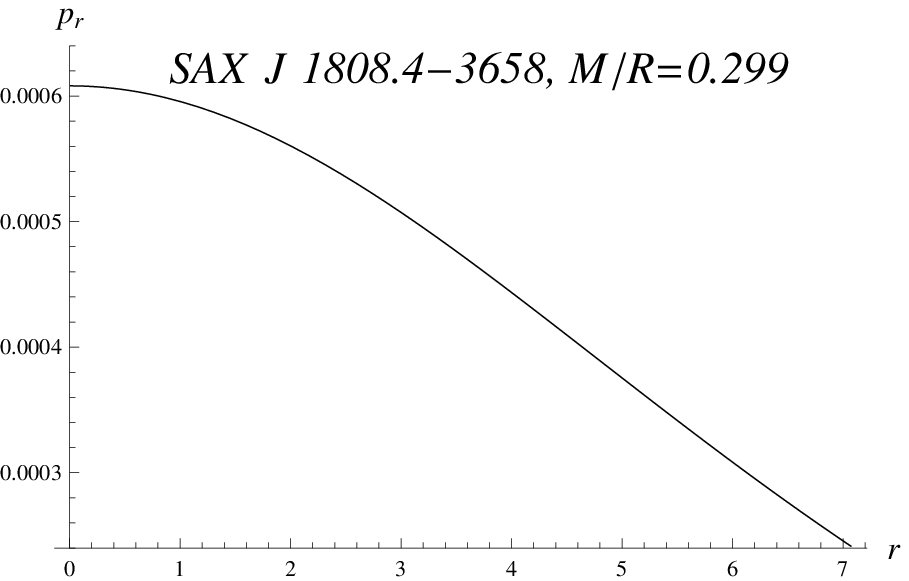, width=.36\linewidth,
height=1.4in}\epsfig{file=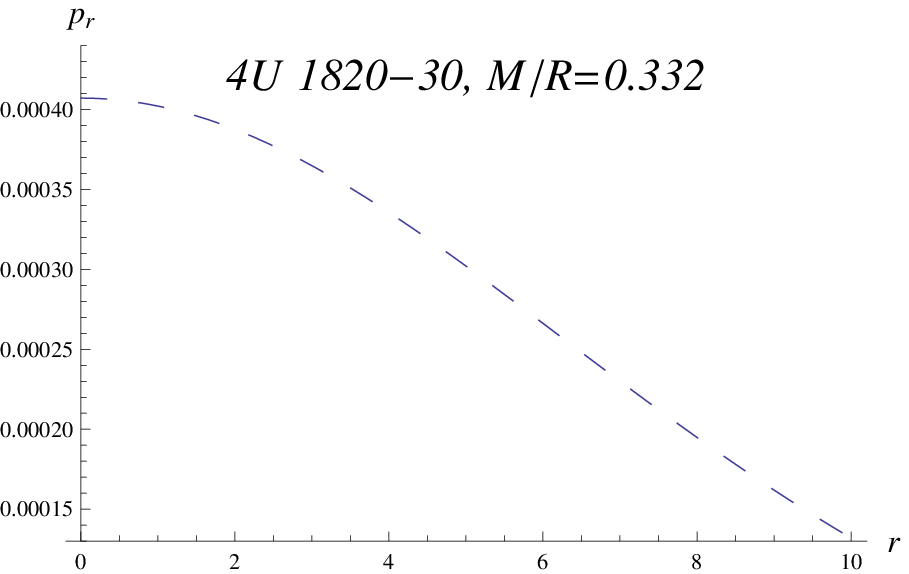, width=.34\linewidth,
height=1.4in}\caption{Evolution of radial pressure $p_r$ versus
$r(km)$ at the stellar interior of strange star candidates.}
\end{figure}
\begin{figure}
\centering \epsfig{file=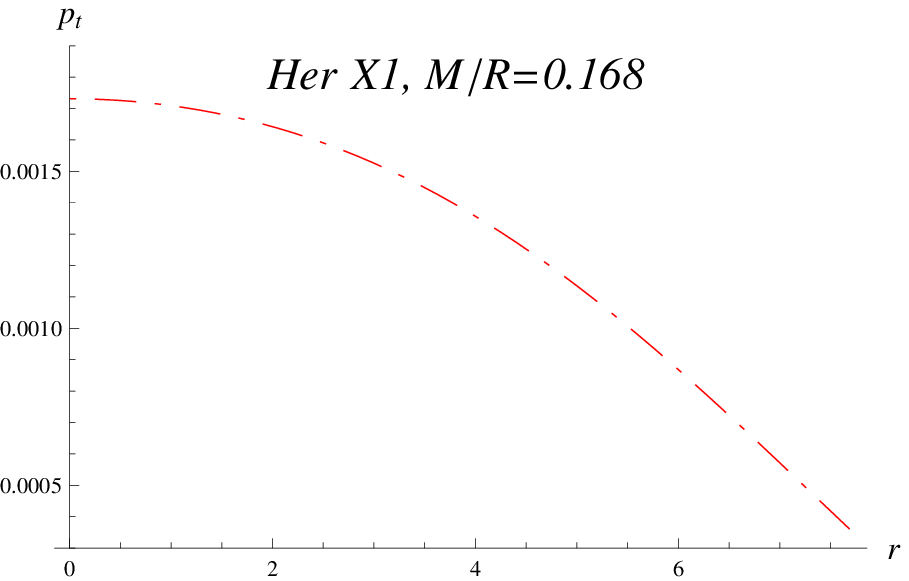, width=.34\linewidth,
height=1.4in}\epsfig{file=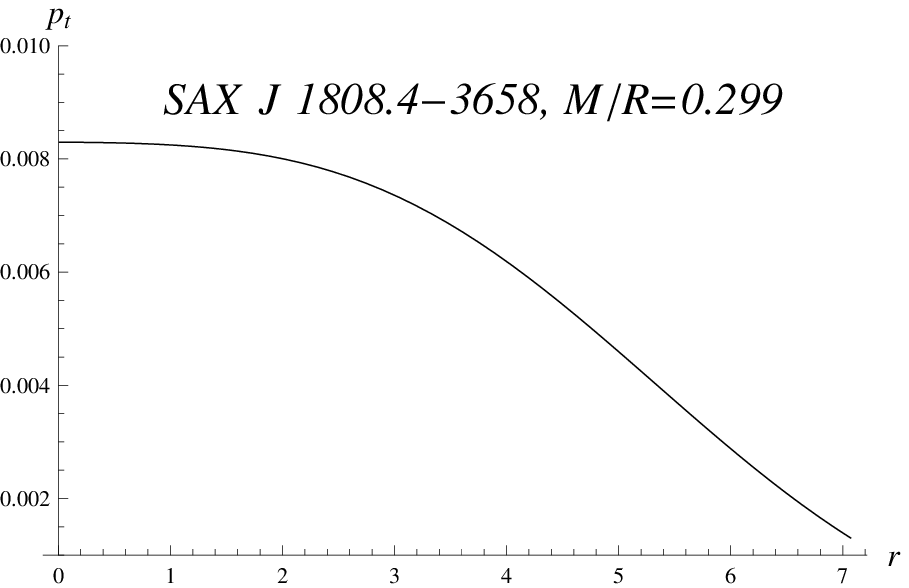, width=.36\linewidth,
height=1.4in}\epsfig{file=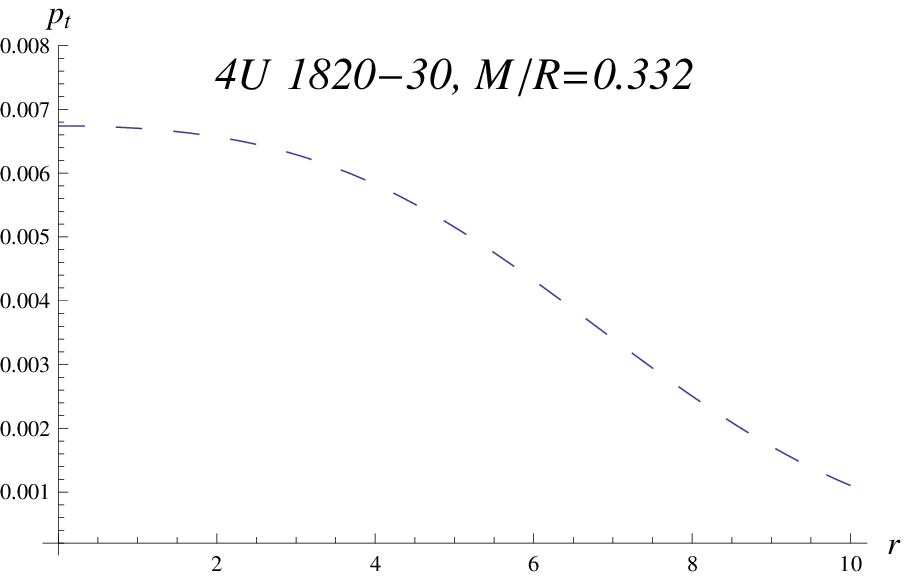, width=.34\linewidth,
height=1.4in}\caption{Evolution of transverse pressure $p_t$
versus $r(km)$ at the stellar interior of strange star candidates.}
\end{figure}
\begin{figure}
\centering \epsfig{file=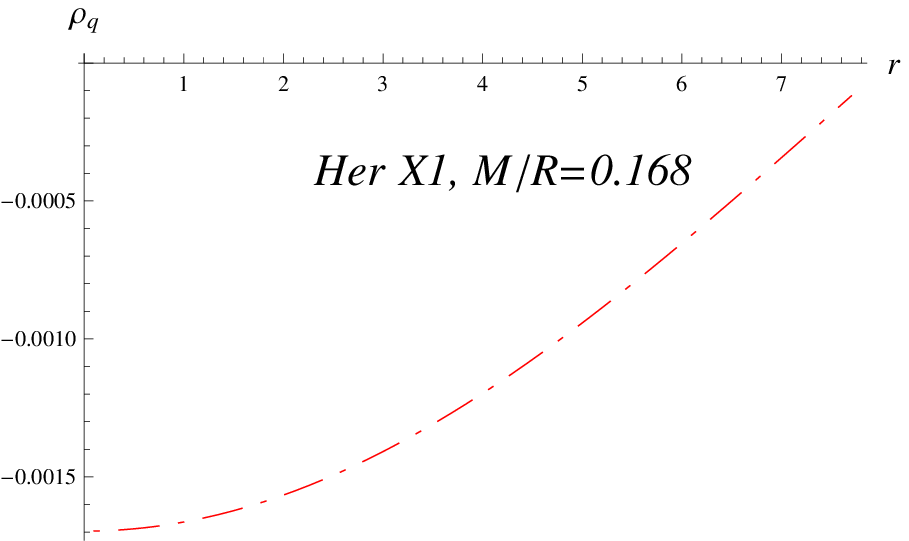, width=.34\linewidth,
height=1.4in}\epsfig{file=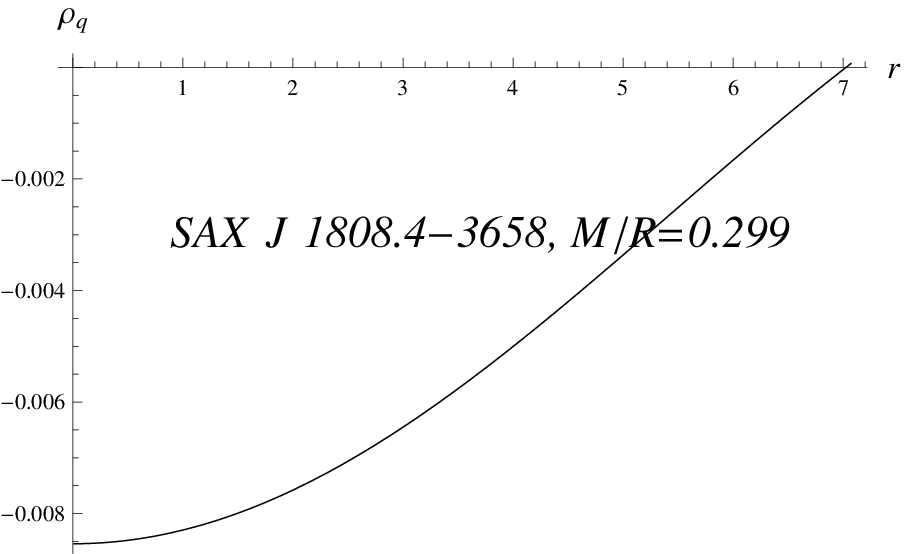, width=.36\linewidth,
height=1.4in}\epsfig{file=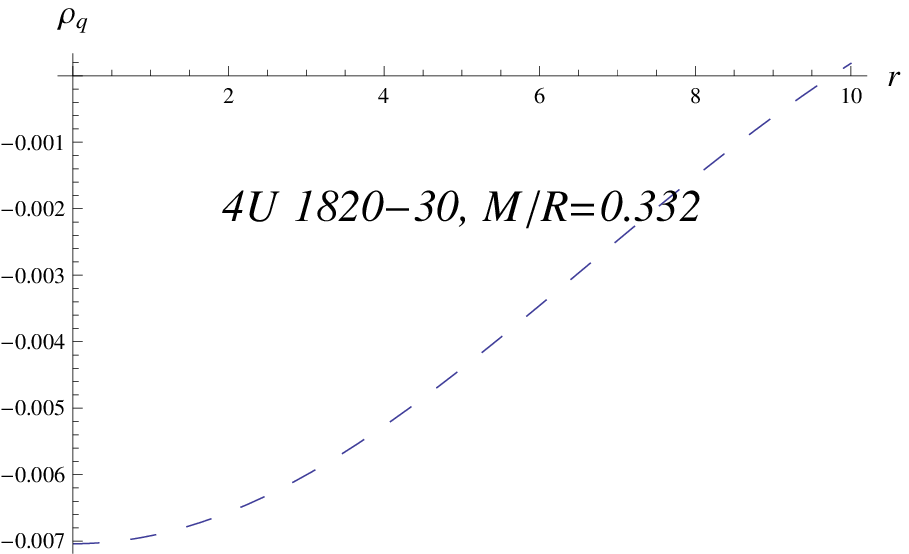, width=.34\linewidth,
height=1.4in}\caption{Evolution of transverse pressure $p_t$ versus
$r(km)$ at the stellar interior of strange star candidates.}
\end{figure}

\section{Physical Analysis}

Here, we discuss some physical conditions which are necessary for
the interior solution. In the following, we present the anisotropic
behavior and stability conditions.

\subsection{Anisotropic Constraints}

Taking derivatives of equations (\ref{15b}) and (\ref{15d}) with
respect to radial coordinate, we have
\begin{eqnarray}\nonumber
\frac{d\rho}{dr}&=&-\frac{1}{r^5(1+\alpha)}4e^{-2Ar^2}\left\{-24\lambda+4r^2
(-9A-6A^2r^2-2(10A^3-11A^2B\right.\\\nonumber&-&\left.AB^2+B^3)r^4+2A(6A^3-17A^2B+4AB^2+3B^3)r^6
+2A^2(A-B)\right.\\\label{18}&\times&\left.B(3A+B)r^8)\lambda+e^{Ar^2}(A(A+B)r^6+12(2+Ar^2)\lambda)\right\},\\\nonumber
\frac{dp_r}{dr}&=&-\frac{1}{r^5(1+\alpha)}4e^{-2Ar^2}\alpha\left\{-24\lambda+4r^2(-9A
-6A^2r^2-2(10A^3-11A^2B\right.\\\nonumber&-&\left.AB^2+B^3)r^4+2A(6A^3-17A^2B+4AB^2+3B^3)r^6+2A^2(A-B)\right.\\\label{19}&-&\left.B
(3A+B)r^8)\lambda+e^{Ar^2}(A(A+B)r^6+12(2+Ar^2)\lambda)\right\}.
\end{eqnarray}
Similarly, one can find the second derivatives of $\rho$ and $p_r$. We
present the evolution of $\frac{d\rho}{dr}$ and $\frac{dp_r}{dr}$ in Figures
\textbf{5} and \textbf{6} which show that $\frac{d\rho}{dr}<0$ and
$\frac{dp_r}{dr}<0$.
\begin{figure}
\centering \epsfig{file=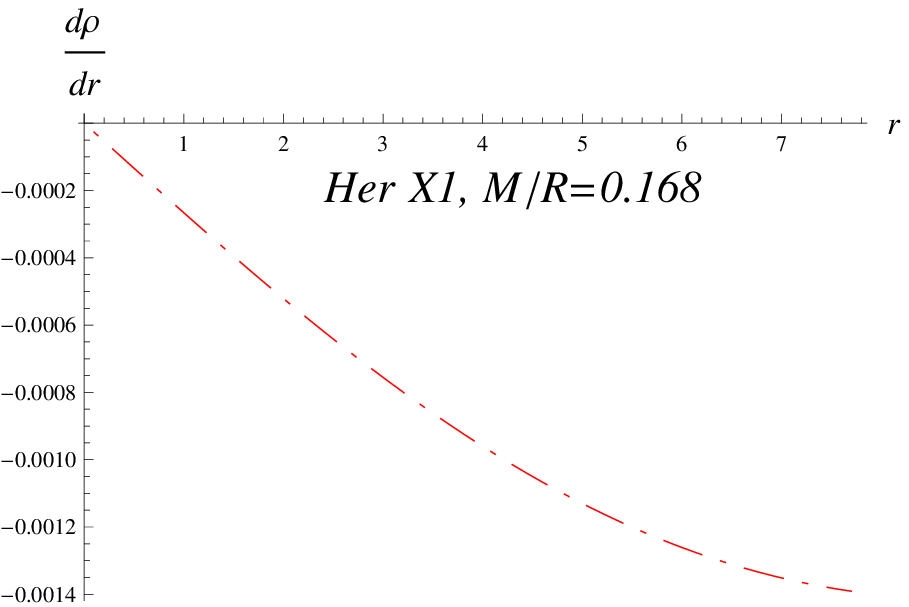, width=.34\linewidth,
height=1.4in}\epsfig{file=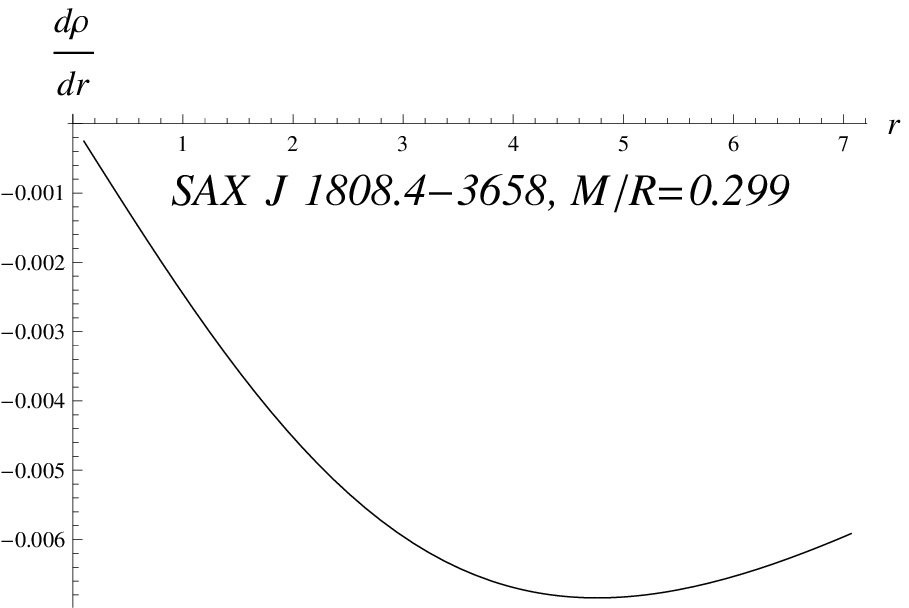, width=.36\linewidth,
height=1.4in}\epsfig{file=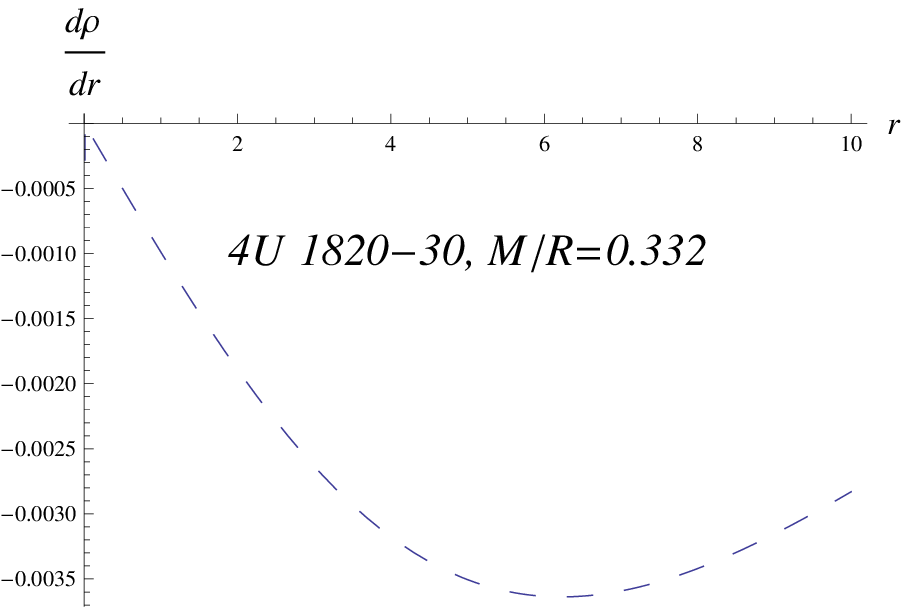, width=.34\linewidth,
height=1.4in}\caption{Evolution of $\frac{d\rho}{dr}$ versus $r(km)$
at the stellar interior of strange star candidates.}
\end{figure}
\begin{figure}
\centering \epsfig{file=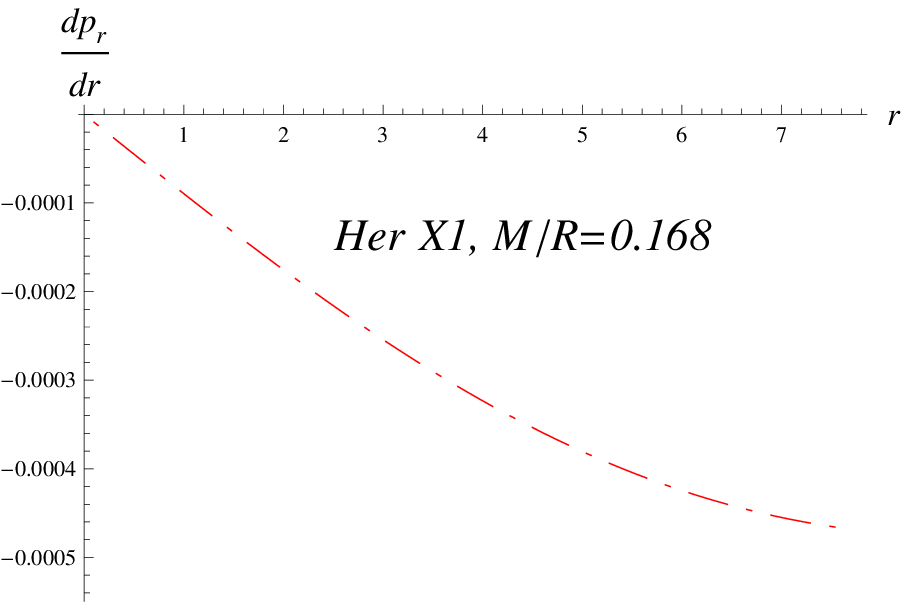, width=.34\linewidth,
height=1.4in}\epsfig{file=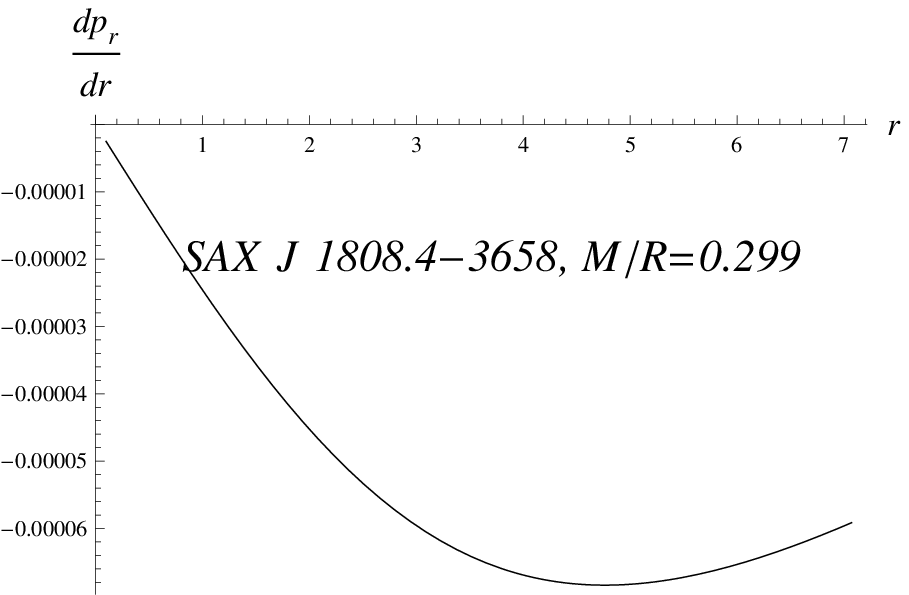, width=.36\linewidth,
height=1.4in}\epsfig{file=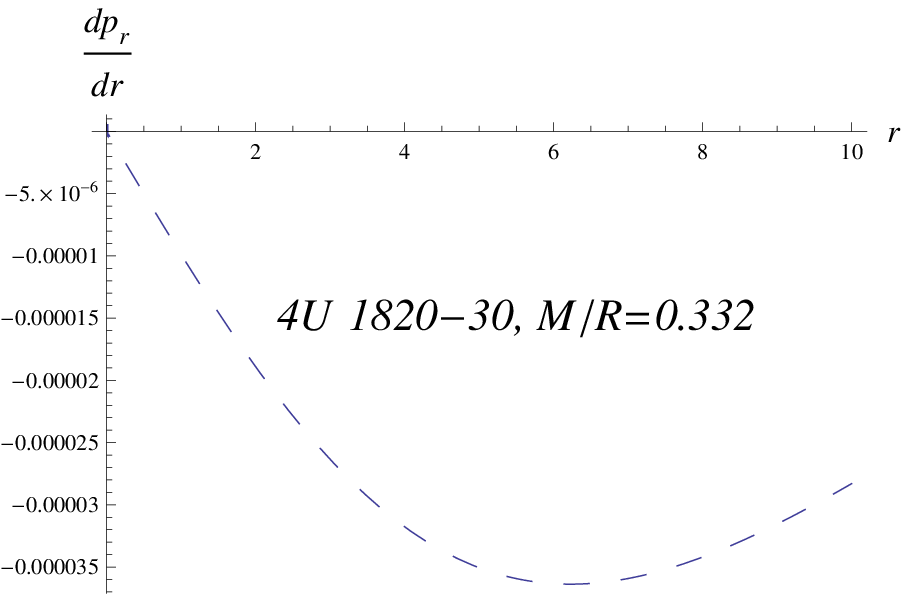, width=.34\linewidth,
height=1.4in}\caption{Evolution of $\frac{p_r}{dr}$ versus $r(km)$
at the stellar interior of strange star candidates.}
\end{figure}

One can explore the behavior of derivatives of $\rho$ and $p_r$ at
center $r=0$ of compact star and it can be seen that
\begin{eqnarray}\nonumber
\frac{d\rho}{dr}=0, \quad \quad \frac{dp_r}{dr}=0,\\\label{20}
\frac{d^2\rho}{dr^2}<0, \quad \quad \frac{d^2p_r}{dr^2}<0.
\end{eqnarray}
This indicate the maximality of central density and radial pressure.
Hence $\rho$ and $p_r$ attain maximum values at $r=0$ and functional
values decreases with the increase in $r$ as shown in Figures
\textbf{1-2}. From Eqs.(\ref{16}) and (\ref{17}), we have
$\omega_r>0$ and $\omega_t<1$ as shown in Figure \textbf{7} for
different strange stars.
\begin{figure}
\centering \epsfig{file=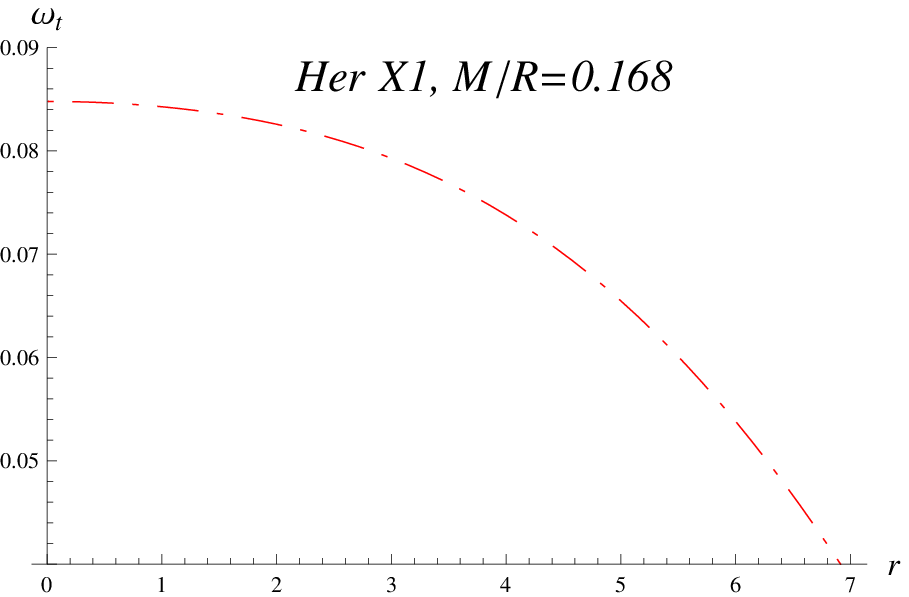, width=.34\linewidth,
height=1.4in}\epsfig{file=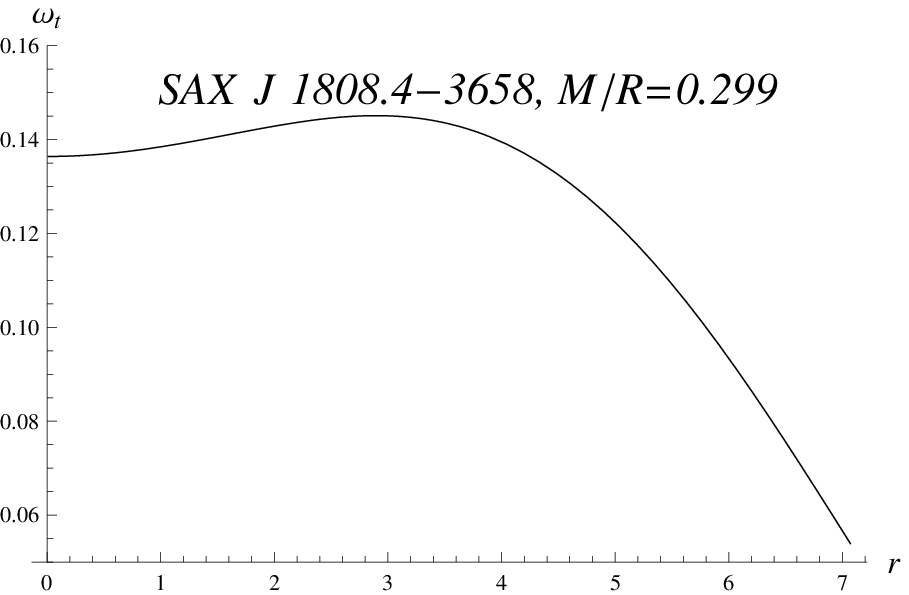, width=.36\linewidth,
height=1.4in}\epsfig{file=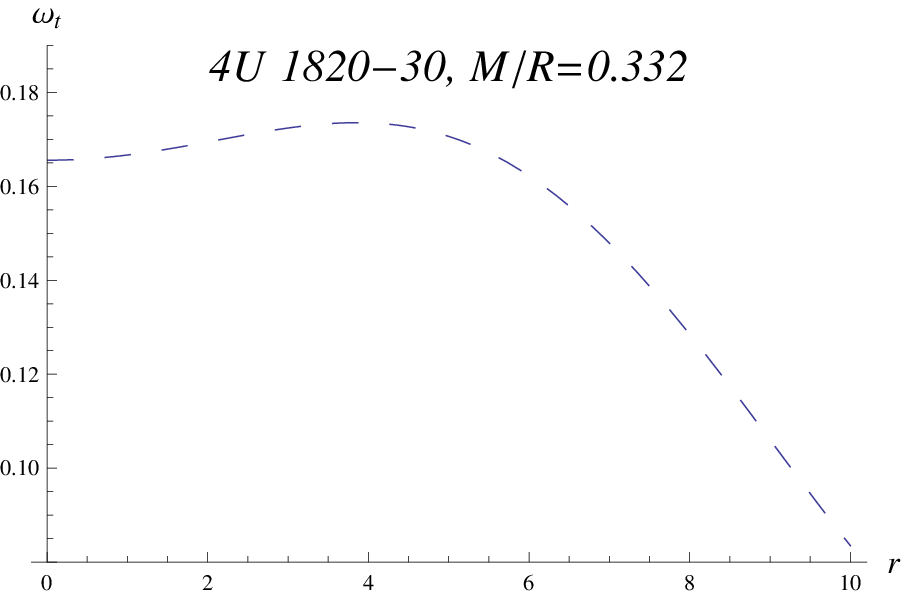, width=.34\linewidth,
height=1.4in}\caption{Evolution of of EoS parameter $\omega_t$
versus $r(km)$ at the stellar interior of strange star candidates.}
\end{figure}

The measure of anisotropy parameter $\Delta=\frac{2}{r}(p_t-p_r)$ in
this case is given by
\begin{eqnarray}\nonumber
\Delta&=&\frac{1}{r^5(1+\alpha)}e^{-2Ar^2}\left\{-e^{2Ar^2}(1+\alpha)(\lambda(2-6\omega_q)
+r^2(1+3\omega_q))+e^{Ar^2}(-2\right.\\\nonumber&\times&\left.(A-B)Br^6(1+\alpha)+r^2(1+\alpha)(1+24(-A+B)\lambda+3\omega_q)
-12\lambda(1\right.\\\nonumber&+&\left.7\alpha+3(-1+\alpha)\omega_q)-2r^4(A-3B(1+\omega_q)+A\alpha(4+3\omega_q)))
-2\lambda(-7\right.\\\nonumber&-&\left.43\alpha+12Br^2(1+\alpha)+3(7-5\alpha)\omega_q-3B^2r^4(11+3\alpha)(1+\omega_q)
+B^4r^8\right.\\\nonumber&\times&\left.(1+\alpha)(-1+3\omega_q)+6B^3r^6(-3+\alpha+(-1+3\alpha)\omega_q)+12A^3r^6(2\right.\\\nonumber&+&\left.Br^2)(2+
\alpha(5+3\omega_q))+2Ar^2(-2(5+14\alpha+3(-2+\alpha)\omega_q)+B^3r^6(7\right.\\\nonumber&+&\left.\alpha-3(-1+\alpha)\omega_q)
+6Br^2(9+13\alpha+4\times(2+3\alpha)\omega_q)+B^2r^4(59+71\alpha\right.\\\nonumber&+&\left.3(9+13\alpha)\omega_q))-A^2r^4(4(17
+38\alpha+(9+30\alpha)\omega_q)+B^2r^4\times(34+\alpha(61\right.\\\label{21}&+&\left.33\omega_q))+Br^2(r^2(3+9\omega_q)+4(40+82\alpha
+(9+51\alpha)\omega_q))))\right\}.
\end{eqnarray}
Figure \textbf{8} shows the evolution of $\Delta$ for the different
strange stars. It can be seen that $\Delta>0$, which implies that it
is directed outward and repulsive force exists for these strange
star models.
\begin{figure}
\centering \epsfig{file=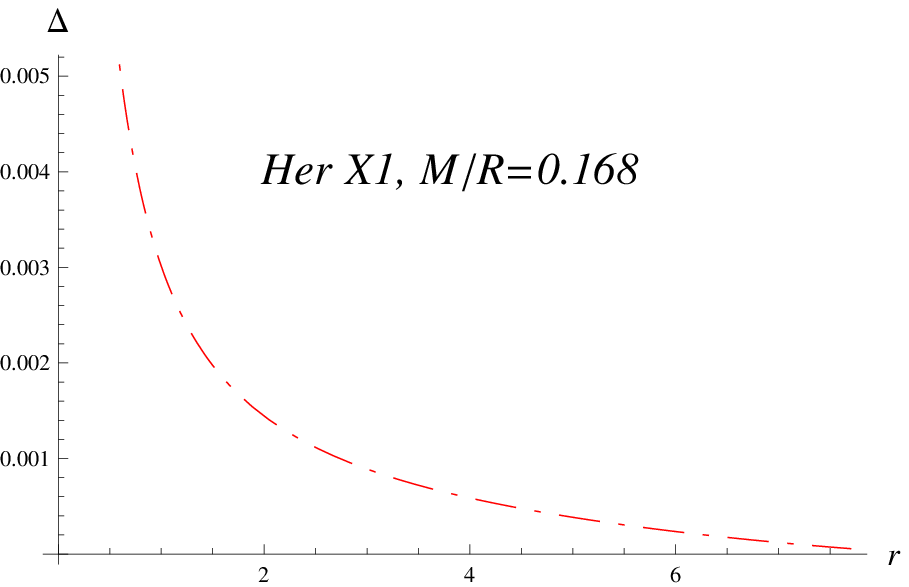, width=.34\linewidth,
height=1.4in}\epsfig{file=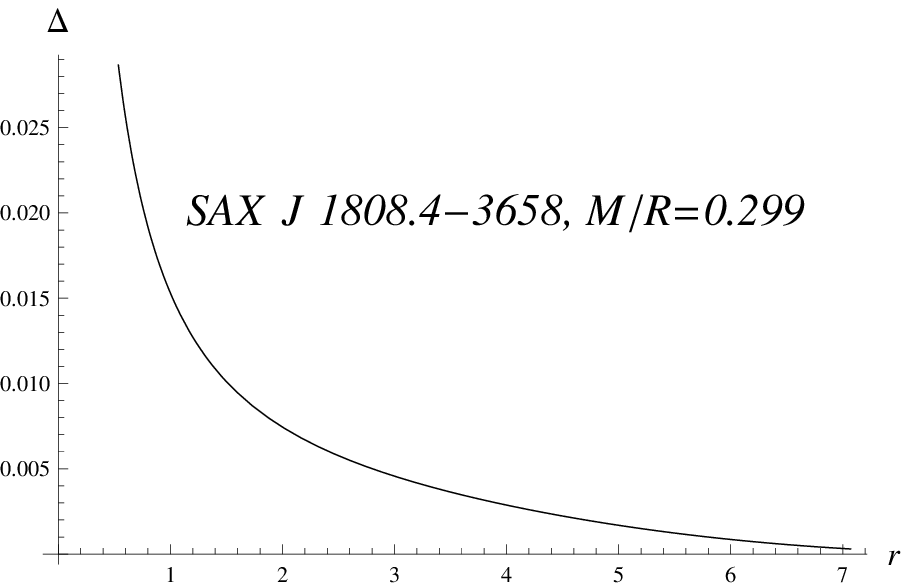, width=.36\linewidth,
height=1.4in}\epsfig{file=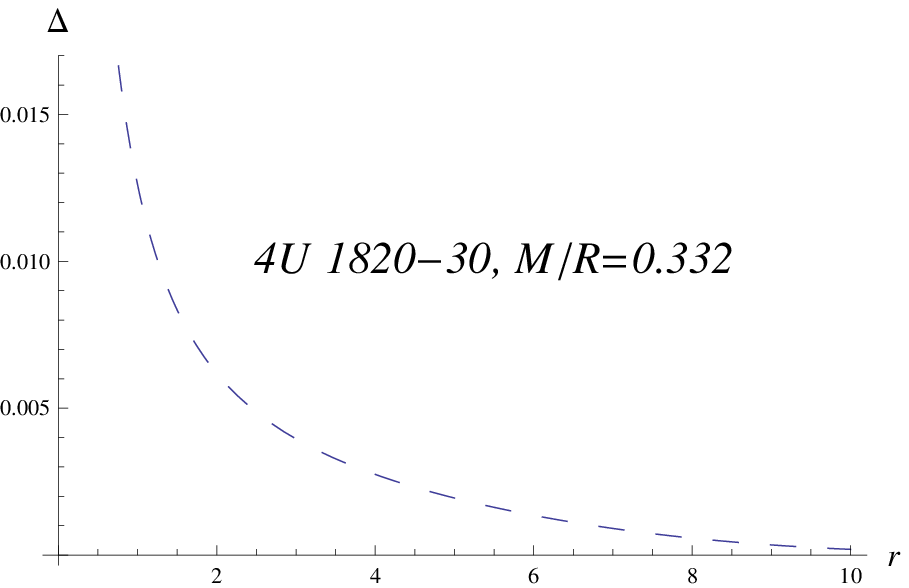, width=.34\linewidth,
height=1.4in}\caption{Variation of anisotropy measurement versus
radial coordinate $r(km)$ at the stellar interior of strange star
candidates.}
\end{figure}

\subsection{Matching Conditions}

Recently, Goswami et al.(2014), have proved that {extra matching
conditions that arise in the modified gravity imposes strong
constraints on the stellar structure and thermodynamic properties.
They showed that these constraints are non-physical}. According to
these authors, Schwarzschild solution is best choice in the exterior
region for matching conditions. It means there does not exist
general vacuum solution in $f(R)$ gravity as Schwarzschild solution
in GR. Using this philosophy a lot of work in $f(R)$ gravity
(Ganguly 2014, Ifra and Zubair 2015, Ifra et al.2015) has been done
by taking the exterior solution as Schwarzschild or Vaidya metric.

Here, we match the interior metric (\ref{4}) to the vacuum exterior
spherically symmetric metric given by
\begin{equation}\label{21}
 ds^2=-\left(1-\frac{2M}{r}\right)dt^2+\left(1-
 \frac{2M}{r}\right)^{-1}dr^2+r^2d\theta^2+r^2 sin^2{\theta}d\varphi^2,
\end{equation}
At the boundary surface $r=R$ continuity of the metric functions
$g_{tt}$, $g_{rr}$ and $\frac{\partial g_{tt}}{\partial r}$ yield,
\begin{eqnarray}\label{22}
  g_{tt}^-=g_{tt}^+,~~~~~
   g_{rr}^-=g_{rr}^+,~~~~~
   \frac{\partial g_{tt}^-}{\partial r}=\frac{\partial g_{tt}^+}{\partial r},
  \end{eqnarray}
where $-$ and $+$, correspond to interior and exterior solutions.
From the interior and exterior metrics, we get
 \begin{eqnarray}\label{23}
  A&=&-\frac{1}{R^2}ln\left(1-\frac{2M}{R}\right),\\\label{24}
 B&=&\frac{M}{R^3}{{\left(1-\frac{2M}{R}\right)}^{-1}},\\\label{24a}
 C&=&ln\left(1-\frac{2M}{R}\right)-\frac{M}{R}{{\left(1-\frac{2M}{R}\right)}^{-1}}.
\end{eqnarray}
For the given values of $M$ and $R$ (Li 1999, Lattimer 2014) of the
compact stars, the constants $A$ and $B$ are given in the table
\textbf{1}.
\begin{table}[ht]
\caption{Values of constants for given Masses and Radii of Stars}
\begin{center}
\begin{tabular}{|c|c|c|c|c|c|}
\hline {Strange Quark Star}&  \textbf{ $M$} & \textbf{$R(km)$} &
\textbf{ $\frac{M}{R}$} &\textbf{ $A(km ^{-2})$}& \textbf{$B(km
^{-2})$}
\\\hline  Her X-1& 0.88$M_\odot$& 7.7&0.168&0.006906276428 &
$0.004267364618$
\\\hline SAX J 1808.4-3658& 1.435$M_\odot$& 7.07&0.299& 0.01823156974 &
$0.01488011569$
\\\hline 4U 1820-30&2.25$M_\odot$& 10.0 &0.332&0.01090644119 &
$0.009880952381$
\\\hline
\end{tabular}
\end{center}
\end{table}

\subsection{Energy Conditions}

Energy constraints have many useful applications in GR as well as in modified
gravity theories (discussion of various cosmological geometries). These
inequalities are firstly formulated in the context of GR for the derivation
of some general results involving strong gravitational fields. In GR, four
types of energy constraints are formulated using a well-known geometrical
results refereed as Raychaudhuri equation (explaining the dynamics of matter
bits). These constraints are labeled as WEC, DEC, NEC and SEC. For an
anisotropic fluid (\ref{5}), these are defined as
\begin{eqnarray}\nonumber
\textbf{NEC}:\quad&&\rho+p_r\geq0, \quad \rho+p_t\geq0,\\\nonumber
\textbf{WEC}:\quad&&\rho\geq0, \quad \rho+p_r\geq0, \quad
\rho+p_t\geq0,\\\nonumber \textbf{SEC}:\quad&&\rho+p_r\geq0, \quad
\rho+p_t\geq0, \quad \rho+p_r+2p_t\geq0,\\\nonumber
\textbf{DEC}:\quad&&\rho>|p_r|, \quad \rho>|p_t|.
\end{eqnarray}
We find that our model satisfies these conditions for specific
values of mass and radius which helps to find the unknown parameters
for different strange stars. Here, we present the evolution of these
conditions for strange star Her X-1 as shown in Figure \textbf{9}.
It can be seen that energy conditions are satisfied for our model.
\begin{figure}
\centering \epsfig{file=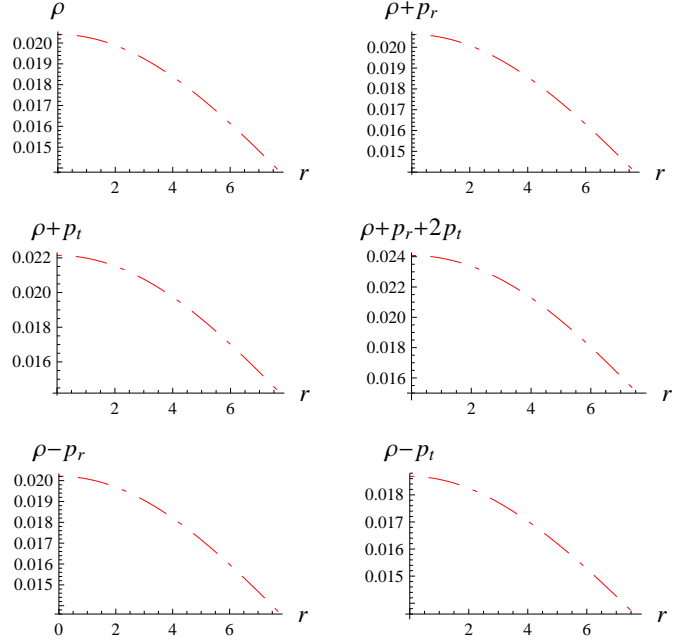}\caption{Evolution of energy
constraints at the stellar interior of strange star Her X-1.}
\end{figure}

\subsection{Stability Analysis}

In this section, we discuss the stability of quintessence star
models in $f(R)$ theory. To analyze the stability of our model we
calculate the radial and transverse speeds as
\begin{figure}
\centering \epsfig{file=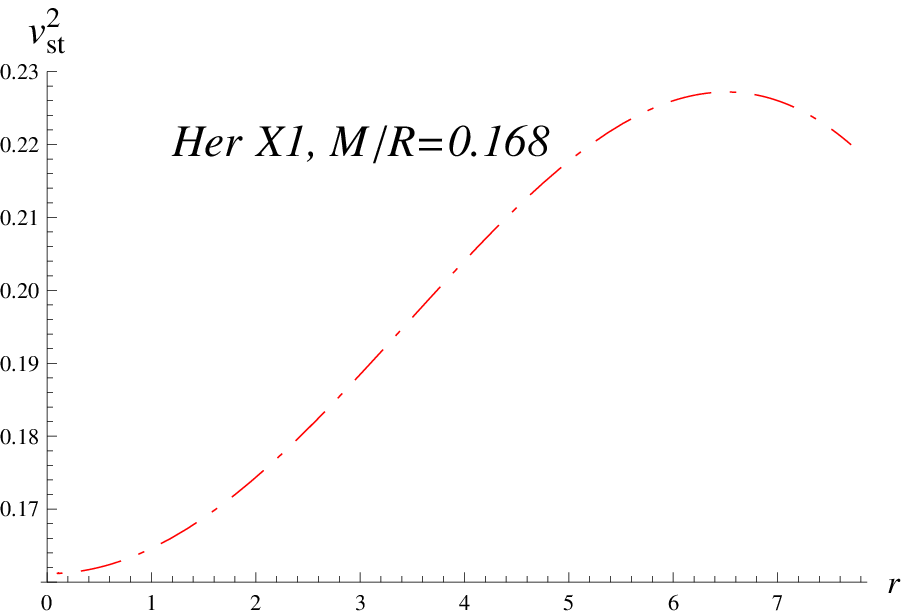, width=.34\linewidth,
height=1.4in}\epsfig{file=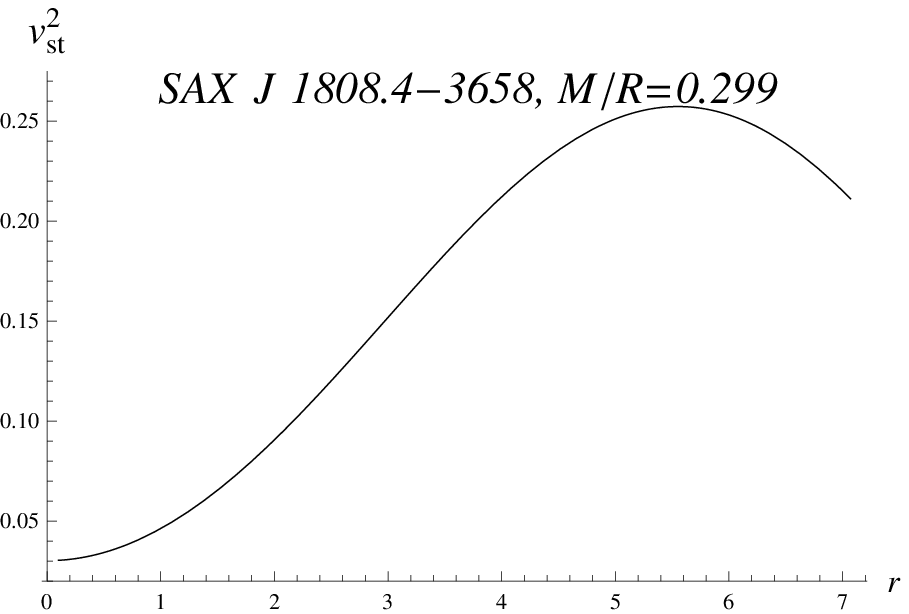, width=.36\linewidth,
height=1.4in}\epsfig{file=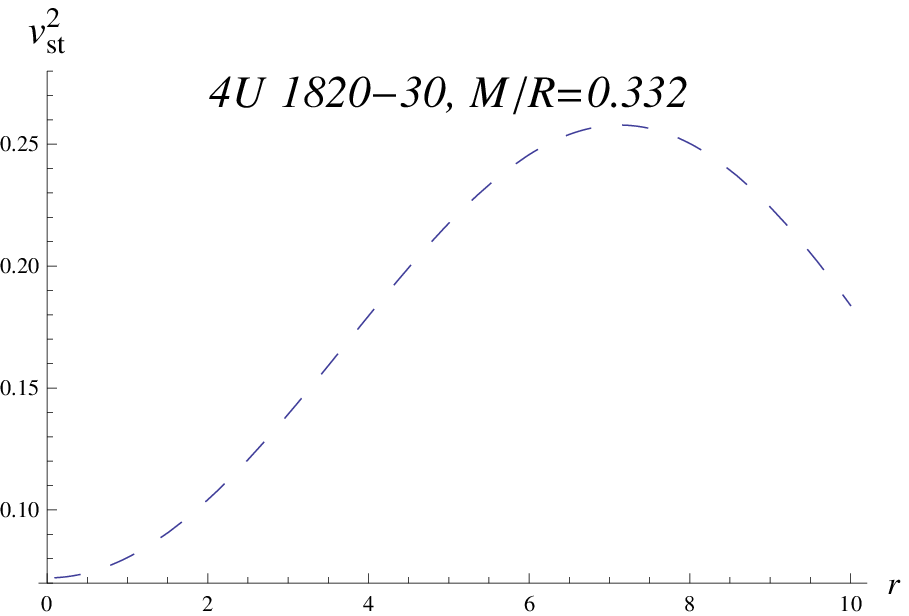, width=.34\linewidth,
height=1.4in}\caption{Variation of $v^2_{st}$ at the stellar
interior of strange star candidates.}
\end{figure}
\begin{eqnarray}\label{22}
v^2_{sr}&=&\alpha, \\\nonumber
v^2_{st}&=&\left\{-e^{2Ar^2}(1+\alpha)(4\lambda(1-3\omega_q)+r^2(1+3\omega_q))-e^{Ar^2}(2A(A-B)Br^8(1
\right.\\\nonumber&+&\left.\alpha)+Ar^4(1+\alpha)(-1+24(A-B)\lambda-3\omega_q)+24\lambda(1+3\alpha+3(-1\right.\\\nonumber&+&\left.\alpha)\omega_q)
+2r^6(B^2(1+\alpha)-AB(4+3\alpha+3\omega_q)+A^2(1+\alpha(2+3\omega_q)))\right.\\\nonumber&+&\left.r^2((1+\alpha)(-1-3\omega_q)
+12\lambda(-2B(1+\alpha)+A(3+5\alpha+3(-1+\alpha)\right.\\\nonumber&\times&\left.\omega_q))))-4\lambda(-7-19\alpha+6Br^2(1+\alpha)+3(7-5\alpha)
\omega_q-B^4r^8(1+\alpha)\right.\\\nonumber&\times&\left.(-1+3\omega_q)+B^3r^6(9+5\alpha+(3-9\alpha)\omega_q)+12A^4r^8(2+Br^2)(2
+3\alpha\right.\\\nonumber&\times&\left.(1+\omega_q))+Ar^2(-17+12Br^2(1+\alpha)+33\omega_q+B^4r^8(1+\alpha)(-1+3\omega_q)\right.\\\nonumber&-&\left.7\alpha(5
+3\omega_q)-4B^2r^4(23+22\alpha+3(5+4\alpha)\omega_q)+4B^3r^6(-8-5\alpha\right.\\\nonumber&+&\left.(-3+6\alpha)\omega_q))
+A^2r^4(-4(5+8\alpha+3(-2+\alpha)\omega_q)+2B^3r^6(7+5\alpha\right.\\\nonumber&-&\left.3(-1+\alpha)\omega_q)+B^2r^4(19(8
+9\alpha)+3(18+37\alpha)\omega_q)+Br^2(r^2(3+9\omega_q)\right.\\\nonumber&+&\left.2(94+116\alpha+3(19+41\alpha)\omega_q)))-A^3r^6
(4(23+33\alpha+(9+39\alpha)\omega_q)\right.\\\nonumber&+&\left.B^2r^4(34+\alpha(45+33\omega_q))+Br^2(r^2(3+9\omega_q)+4
(46+63\alpha+(9+60\alpha)\right.\\\nonumber&\times&\left.\omega_q))))\right\}/\left\{4(-24\lambda+4r^2(-9A-6A^2r^2-2(10A^3-11A^2B-AB^2+B^3)r^4
\right.\\\nonumber&+&\left.2A(6A^3-17A^2B+4AB^2+3B^3)r^6+2A^2(A-B)B(3A+B)r^8)\lambda\right.\\\label{23}&+&\left.e^{Ar^2}(A(A+B)r^6+12(2+Ar^2)\lambda))\right\}.
\end{eqnarray}

In the past, Herrera and his collaborators (Herrera 1992, Chan et
al.1993, Di Prisco 1997)have developed a new technique to explore
the potentially unstable matter configuration and introduced the
concept of cracking. One can analyze the potentially stable and
unstable regions regions depending on the difference of sound
speeds, the region for which radial sound speed is greater than the
transverse sound speed is said to be potentially stable. We find
that radial sound speed is constant (Eq.(\ref{22})) and plot the
transverse sound speed in Figure \textbf{10}. It can be seen that
$v^2_{st}$ satisfy the relation of stable matter configuration
$0\leq{v}^2_{st}\leq1$. This variation is further confirmed in
Figure \textbf{11}, where difference of the two sound speeds, i.e.,
$v^2_{st}-v^2_{sr}$ retain similar sign within the specific
configuration and it satisfies the inequality
$|v^2_{st}-v^2_{sr}|\leq1$. Thus, our proposed strange star model is
stable.
\begin{figure}
\centering \epsfig{file=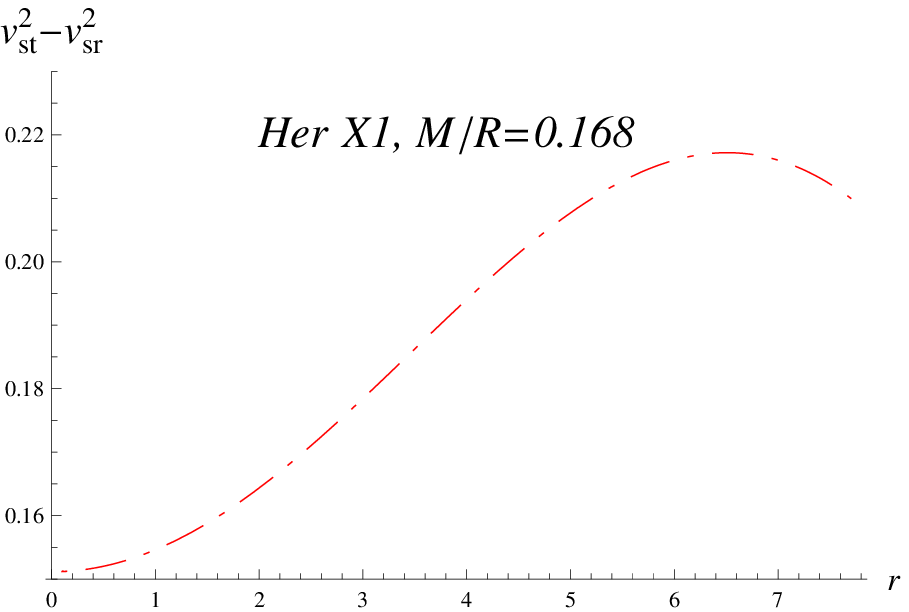, width=.34\linewidth,
height=1.4in}\epsfig{file=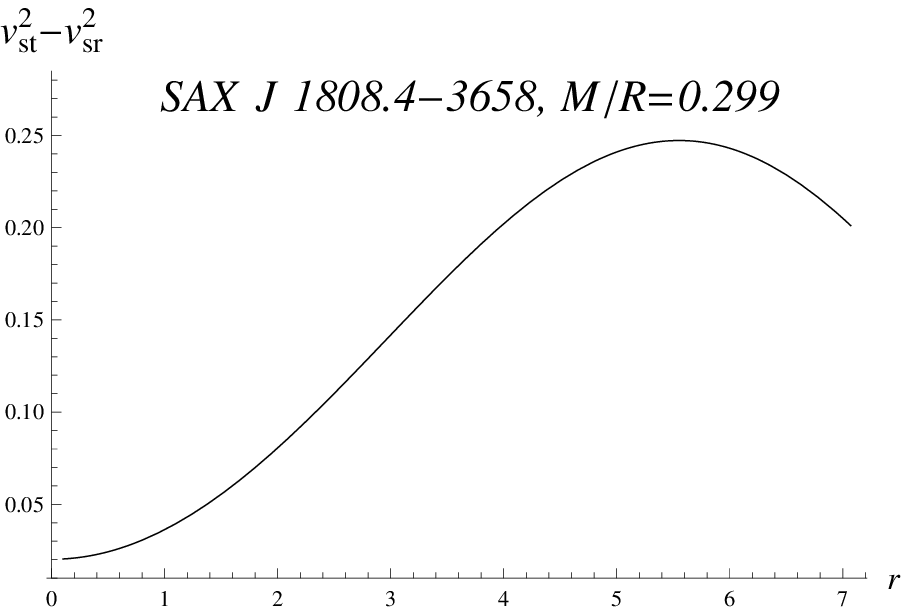, width=.36\linewidth,
height=1.4in}\epsfig{file=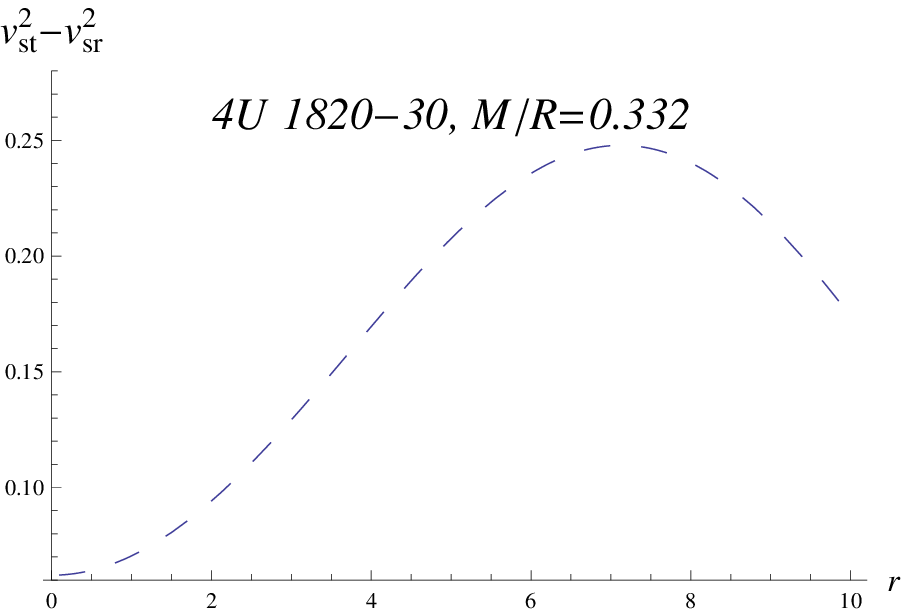, width=.34\linewidth,
height=1.4in}\caption{Variation of $v^2_{st}-v^2_{sr}$ at the stellar
interior of strange star candidates.}
\end{figure}

\subsection{Surface Redshift}

The compactness of the star is given by
\begin{eqnarray}\nonumber
u&=&\frac{M(R)}{R}=\frac{1}{8(AR^2)^{5/2}(1+\alpha)}R^2e^{-2AR^2}\pi\left\{-4\sqrt{A}(9B^3R^2\lambda
+48A^4R^4(2\right.\\\nonumber&+&\left.BR^2)\lambda-4A^3R^2(10+BR^2(41+8BR^2))\lambda+2ABR^2(4e^{AR^2}+B(13\right.\\\nonumber&+&\left.6BR^2)\lambda)+A^2(-(192
+BR^2(59+8BR^2(1+2BR^2)))\lambda+8e^{AR^2}(R^2\right.\\\nonumber&+&\left.24\lambda)))-e^{2AR^2}\sqrt{\pi}R(16A(-A-B+48A^2\lambda)\text{Erf}
[\sqrt{A}R]-\sqrt{2}(536A^3\right.\\\label{24}&-&\left.59A^2B+26AB^2+9B^3)\lambda\text{Erf}[\sqrt{2}\sqrt{A}R])\right\}.
\end{eqnarray}
The surface redshift ($Z_s$) corresponding to compactness (\ref{24}) is given
by
\begin{eqnarray}\nonumber
{1+Z_s}&=&(1-2u)^{-1/2}=\{1-\frac{R^2e^{-2AR^2}\pi}{4(AR^2)^{5/2}(1+\alpha)}\left\{-4\sqrt{A}(9B^3R^2\lambda
+48A^4R^4(2\right.\\\nonumber&+&\left.BR^2)\lambda-4A^3R^2(10+BR^2(41+8BR^2))\lambda+2ABR^2(4e^{AR^2}+B(13\right.\\\nonumber&+&\left.6BR^2)\lambda)+A^2(-(192
+BR^2(59+8BR^2(1+2BR^2)))\lambda+8e^{AR^2}(R^2\right.\\\nonumber&+&\left.24\lambda)))-e^{2AR^2}\sqrt{\pi}R(16A(-A-B+48A^2\lambda)\text{Erf}
[\sqrt{A}R]-\sqrt{2}(536A^3\right.\\\label{25}&-&\left.59A^2B+26AB^2+9B^3)\lambda\text{Erf}[\sqrt{2}\sqrt{A}R])\right\}\}^{-1/2}.
\end{eqnarray}
In Figure \textbf{12}, we show the evolution of redshift for the strange star
Her X-1.
\begin{figure}
\centering \epsfig{file=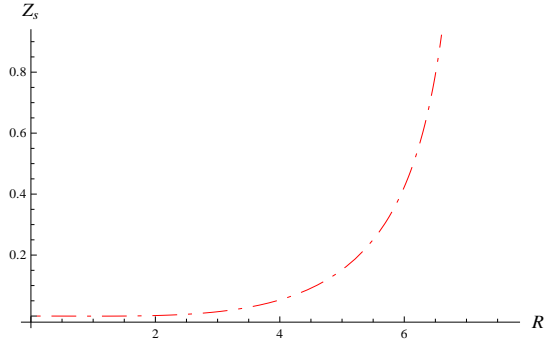, width=.52\linewidth,
height=1.8in}\caption{Evolution of redshift of strange star Her
X-1.}
\end{figure}\\\\
\section{Conclusion}
The current cosmological observations imply that there are two
phases of accelerated expansion in our present model of universe:
cosmic inflation in the past era of the universe and acceleration in
the present expansion of the universe. The investigation of current
cosmic expansion and nature of dark energy has been widely accepted
among the huge community of the scientists. For this purpose,
several attempts have been made for the different strategies to
modify the General Relativity. The $f(R)$ gravity is one of the
modifications of General Relativity.

The current paper deals with the investigation of analytical models
of quintessence compact stars with the anisotropic gravitating
static source in the framework of $f(G)$ gravity. To this end, we
have choosen the Starobinsky model of the form
$f(R)=R+\lambda{R}^2,$ further the stars are assumed as anisotropic
in their internal structure. The analytic solution in $f(R)$ gravity
have found by matching the interior spacetime with the well-known
exterior vacuum spacetime. This matching is suitable in this case as
$f(T)$ and GR both involve second order derivative terms in the
equations of motion and we have continuity of metric coefficients up
to first order derivatives. The graphical behavior of the results
exhibit the some prominent properties of the anisotropic
quintessence compact stars in $f(R)$ gravity.

We have evaluated the matter density, radial and transverse
pressures, quintessence energy density and anisotropic parameter of
the model. Using the observational data of
SAXJ1808.4-3658(SS1)(radius=7.07 km),4U1820- 30 (radius=10 km),  Her
X-1 (radius=7.7 km), we have plotted the energy density, pressure
and quintessence density  at center $r=0$ to the boundary of the
corresponding star. All this results have been shown in figure
\textbf{1-4}. The first and second derivatives of density and
pressures shown in figures \textbf{5,6}, indicate that these
quantities have maximum values at the center and minimum values at
boundary. The graphical behavior of quintessence density $\rho_{q}$
does not change in $f(R)$ theory of gravity as compared to GR, but
there occur a deviation in numerical values \textbf{4}. The
constraint on the EoS parameter is given by $0<{\omega}_t<1$ (as
shown in figure \textbf{7}) which is in agreement with normal matter
distribution in $f(R)$ gravity. We have investigated that for our
model $\Delta>0$ (as shown in figure \textbf{8}) and a repulsive
force due to anisotropy results to the formation of more massive
stars. The proposed model satisfy the energy conditions, as an
example we have shown in figure \textbf{9} that these conditions are
satisfied for $Her-X1 (radius=7.7km)$. We have shown that $v
_{st}^{2}<1$ (see figure \textbf{10}) and $v _{st}^{2}>v _{sr}^{2}$
(see figure \textbf{11}), hence our model is potentially stable. The
range of surface redshift $Z_s$ for compact star candidate
$Her-X1(radius=7.7 km)$ is shown in figure \textbf{12}.

\section{Conflict of Interest}

The authors declare that there is no conflict of interest regarding
the publication of this work.\\

\end{document}